\documentclass[11pt,a4paper]{article}
\usepackage{jcappub}

\newcommand{\lcdm}{\mathrm{\Lambda CDM}}
\newcommand{\densm}{\Omega_{\mathrm{m}}}
\newcommand{\densl}{\Omega_{\mathrm{\Lambda}}}

\title{Quadrature algorithms to the luminosity distance with a time-dependent dark energy model}
\author[a]{Nan-Nan Yue,}
\author[a]{De-Zi Liu,}
\author[a]{Xiao-Xing Pei,}
\author[b]{Fang-Fang Zhu,}
\author[a,c]{Tong-Jie Zhang }
\author[a]{and Zhi-Liang Yang}

\affiliation[a]{Department of Astronomy, Beijing Normal University, Beijing 100875,
P.~R.~China}
\affiliation[b]{School of Mathematical Science, Beijing Normal University, Beijing 100875,
P.~R.~China}
\affiliation[c]{Center for High Energy Physics, Peking University, Beijing
100871, P.~R.~China}

\emailAdd{yuenannan@mail.bnu.edu.cn}
\emailAdd{bingzi@mail.bnu.edu.cn}
\emailAdd{peixing624@mail.bnu.edu.cn}
\emailAdd{fiona-90@live.cn}
\emailAdd{tjzhang@bnu.edu.cn}
\emailAdd{zlyang@bnu.edu.cn}

\abstract{
In our previous work \cite{2011MNRAS.412.2685L}, we have proposed two methods for computing the luminosity distance $d_{L}^{\Lambda}$ in $\lcdm$ model. In this paper, two effective quadrature algorithms, known as Romberg Integration and composite Gaussian Quadrature, are presented to calculate the luminosity distance $d_{L}^{CPL}$ in the Chevallier-Polarski-Linder parametrization(CPL) model. By comparing both the efficiency and accuracy of the two algorithms, we find that the second is more promising. Moreover, we develop another strategy adapted for approximating $d_{L}^{\Lambda}$ in flat $\lcdm$ universe. To some extent, our methods can make contributions to the recent numerical stimulation for the investigation of dark energy cosmology.}

\arxivnumber{1109.6388v1}

\begin{document}
\maketitle

 \flushbottom

\section{Introduction}\label{intro}
The computation and numerical evaluation of distances is frequently encountered
in the research of cosmological phenomena.  In practice, it is common to
compute the various cosmological distances as a function of the redshift $z$
under certain cosmological models. Current cosmological observations indicate
that the expansion of universe is accelerating, has a prominent dark energy content $\lcdm \simeq 0.7$
and is spatially flat \cite{2010ApJ...716..712A}.  Further, a Lambda cold dark matter ($\lcdm$) model fits the data well, and is
frequently used as a fiducial or background model.  In the $\lcdm$ model,
various cosmological distances can be expressed in terms of the elliptic
integrals \cite{1997astro.ph..9054E,1999astro.ph..5116H}.

Some works have focused on the luminosity distance $d_L$ in the $\lcdm$
model and derived numerical approximations for the efficient and accurate
evaluation of $d_L(z)$ given the cosmological parameters $\densm$ and $\densl$
\cite{1999ApJS..120...49P,2010MNRAS.406..548W,2010arXiv1012.2670A}.
The computation of $d_L$ is useful in the analysis of distance-redshift relations
of type Ia supernovae, and the approximation for $d_L$ can also be directly used in
the evaluation of other distances, for instance the angular diameter distance $d_{A}$
or the comoving distance $r$ \cite{1999astro.ph..5116H}.

Recent years, Chevallier, Polarski \cite{CP} and Linder \cite{Linder} proposed a simple parametrization of the dark energy
equation of state (known as CPL):
\begin{eqnarray}
\label{DE_EoS}
w(z) = w_{0} + w_{a}\frac{z}{1 + z},
\end{eqnarray}
which is involved in the luminosity distance. The best fit values of $w_{0}$ and $w_{a}$ are -1.58 and 3.30\cite{Linden},
respectively. CPL parametrization is widely applied into both observational and theoretical analysis,
and it has the talent to test the dynamics of many dark energy models. More discussions about CPL model can be seen \cite{{Xin},{Linder},Linden,Dicus,{Linder2}}. In this paper, we just concentrate on the numerical analysis of the complicated integral contained in the luminosity distance of CPL parametrization model. Note that analogous integral may be encountered in many cases,
e.g. the dynamical age of the universe or the angular diameter distance.

For the rest of this paper we will pay our main attention to perform the numerical quadrature algorithms.
Section \ref{CPL_dl} is a brief review of the luminosity distance in the CPL parametrization model.
In section \ref{quadrature}, we present two different quadrature algorithms and compare their efficiency and accuracy based on
the personal computer. Another approximate recipe of the luminosity distance in $\lcdm$ is developed in section \ref{lcdm_dl}.
Finally, we discuss some improvements of the algorithms and possible extensions to other cosmological models briefly.

\section{Luminosity Distance in CPL Model}\label{CPL_dl}
In order to study the different dark energy models, the widely used method is assume an \textit{ad hoc} equation of state
$w(z) = p_{X}/\rho_{X}$ for dark energy and parametrize $w(z)$ \cite{Johri}.
CPL parametrization model was first proposed by M. Chevallier, D. Polarski \cite{CP} and E. V. Linder \cite{Linder},
and the parameterized $w(z)$ can be written as equation \ref{DE_EoS}. Thus the dark energy density $\rho_{X}$ is given by
\begin{eqnarray}
\rho_{X}(z) = \rho_{X}^{0}f(z)
\end{eqnarray}
with
\begin{eqnarray}
\label{f(z)}
f(z) = (1+z)^{3(1 + w_{0} + w_{a})} \exp(-\frac{3w_{a}z}{1+z}).
\end{eqnarray}

For a spatially flat universe ($k = 0$), the Friedmann equation can be expressed as
\begin{eqnarray}
\label{Friedmann}
H^{2}(z) = \frac{8 \pi G}{3}(\rho_{M} + \rho_{X})=H_{0}^{2} E^{2}(z)\nonumber \\
= H_{0}^{2}[\densm(1 + z)^{3} + \Omega_{\Lambda}(z)],
\end{eqnarray}
where $H(z)$ is the Hubble parameter, $\densm$ is the dark matter parameter, $\Omega_{\Lambda}(z)$ represents the time-dependent
dark energy parameter, $E(z)$ is the expansion rate of the universe.

With the continuous equation and Friedmann equation {\ref{Friedmann}}, $\Omega_{\Lambda}(z)$ can be deduced as
\begin{eqnarray}
\Omega_{\Lambda}(z) = \Omega_{\Lambda} f(z),
\end{eqnarray}
where $\Omega_{\Lambda}$ is the dark energy parameter at present time and $f(z)$ is defined as equation {\ref{f(z)}}.
Hence, the luminosity distance in the CPL model can be written in the form
\begin{eqnarray}
\label{dl_CPL}
d_{L}^{CPL} = \frac{c(1+z)}{H_{0}} \int_{0}^{z}\frac{dz'}{E(z')}.
\end{eqnarray}
The expression of the luminosity distance is complicate and there is no critically analytical solution for general parameter choice $(w_{0}, w_{a})$.
Meanwhile, we can take the luminosity distance in $\lcdm$ universe as a special case with $(w_{0}, w_{a}) = (-1.0, 0.0)$.
From this point of view, the $d_{L}^{CPL}$ will degenerate to
\begin{eqnarray}
\label{dl_lcdm}
d_{L}^{\Lambda} = \frac{c(1+z)}{H_{0}}\int_{0}^{z}\frac{dz'}{\sqrt{\densm(1+z')^3 + \densl}}.
\end{eqnarray}

\section{Quadrature Algorithms}\label{quadrature}
Equation {\ref{dl_CPL}} is just a one-dimensional integral, but there are four variables in the integrand,
i.e. $\densm$, $z$, $w_{0}$, $w_{a}$.
Because of the different variations of the three variables $(\densm, w_{0}, w_{a})$, approximating the integral directly seems to be
impossible if we want to obtain desirable accuracy. On the other hand, if we approximate the integrand with multivariables interpolation technique \cite{lancaster} and then integrate the approximate polynomial,
the expression will remain complicate and is hardly to implement in practice.

Instead of developing an analytical approximation, an effective quadrature algorithm may be more helpful.
We present two conventional numerical integration methods, known as Romberg Integration and Gaussian Quadrature {\cite{Krishnamurthy}},
in the following subsections and compare their performances to see which one is more suitable for calculating the luminosity distance.

For simplifying the following description, we just consider the integral in equation {\ref{dl_CPL}}, defined as :
\begin{eqnarray}
\label{fE}
fE = \int_{0}^{z}\frac{dz'}{E(z')}
\end{eqnarray}
with
\begin{eqnarray}
E^{2}(z) = \densm(1 + z)^{3} + \Omega_{\Lambda} (1+z)^{3(1 + w_{0} + w_{a})} \exp(-\frac{3w_{a}z}{1+z}) \nonumber,
\end{eqnarray}
where $\densm + \Omega_{\Lambda} = 1$.

\subsection{Romberg Integration}\label{S_romberg}
Romberg Integration gives preliminary approximations with the Composite Trapezoidal rule and
then applies the Richardson extrapolation process to improve the accuracy.
For each integer $k = 2, 3, 4, ..., n$ and $j = 2, 3, ..., k$, an $O(h_{k}^{2j})$ approximation formula can be written as
\begin{eqnarray}
\int_{a}^{b} f(x)dx = R_{k,j} + O(h_{k}^{2j}),
\end{eqnarray}
where $h_{k} \equiv (b-a)/2^{k-1}$ and the iterative formula $R_{k,j}$ is
\begin{eqnarray}
\label{Romberg}
R_{k,j} = R_{k,j-1} + \frac{R_{k,j-1} - R_{k-1,j-1}}{4^{j-1} - 1}.
\end{eqnarray}
In order to obtain the complete iterative process, we should considerate the case when $j = 1$.
In general, the $R_{k,1}$ is provided by using the trapezoidal approximation, then we have
\begin{eqnarray}
R_{1,1} = \frac{h_{1}}{2}[f(a) + f(b)],
\end{eqnarray}
and
\begin{eqnarray}
R_{k,1} = \frac{1}{2}\left[R_{k-1,1} + h_{k-1} \sum_{i=1}^{2^{k-2}}f(a + (2i - 1)h_{k})\right],
\end{eqnarray}
for $k=2, 3, ..., n$.

\begin{table}[h]
\begin{center}
\caption{\label{Table1} Approximation results for Romberg Integration.}
    \begin{tabular}{lllllllll}
       \hline
       & $R_{1,1}$ &  &  &  & \\
       & $R_{2,1}$ & $R_{2,2}$ &  &  & \\
       & $R_{3,1}$ & $R_{3,2}$ & $R_{3,3}$ &  & \\
       & $R_{4,1}$ & $R_{4,2}$ & $R_{4,3}$ & $R_{4,4}$ & \\
       & $\vdots$ & $\vdots$ & $\vdots$ & $\vdots$ & $\ddots$ & \\
       & $R_{n,1}$ & $R_{n,2}$ & $R_{n,3}$ & $R_{n,4}$ & $\cdots$ & $R_{n,n}$\\
       \hline
     \end{tabular}
\end{center}
\end{table}
The main results generated from the above formulas are listed in Table {\ref{Table1}}.
The Romberg Integration has an additional desirable feature that it allows an entire new row in the table to be calculated
by performing one additional application of the Composite Trapezoidal rule.
Then it uses an averaging of the previously calculated values to obtain the succeeding entries in the row.
The method used to construct a table of this type calculates the entries row by row, that is,
in the order $R_{1,1}$, $R_{2,1}$, $R_{2,2}$, etc. R. L. Burden describes a detailed algorithm in \cite{Burden}.

We can preset an integer $n$ to determine the number of rows. In many cases, however, it is confused to ensure whether
the output is satisfactory or too many entries are unnecessary to generate.
For using the iterative technique  sufficiently and saving the running time,
we can set an error tolerance for the approximation and generate $n$, within some upper bound,
until some consecutive entries agree to within the tolerance.
In this paper, we choose $|R_{n,n} - R_{n,n-1}| < 10^{-6}$ to generate the approximations.

The Romberg Integration method can be used in conjunction with other numerical quadrature formulae to obtain successive improved values.
Further, the method is applicable to a very large class of functions.

\subsection{Composite Gaussian Quadrature}\label{S_Gaussian}
Gaussian Quadrature chooses the points for evaluation in an optimal, rather than equally spaced.
The nodes $r_{1}, r_{2}, ..., r_{n}$ in the interval $[a, b]$ and coefficients $c_{1}, c_{2}, ..., c_{n}$,
are chosen to minimize the expectancy obtained in the approximation
\begin{eqnarray}
\label{Ori_Gauss}
\int_{a}^{b} f(x) dx \approx \sum_{i=1}^{n} c_{i} f(r_{i}).
\end{eqnarray}

With the Legendre polynomials, we can determine the nodes and coefficients easily.
The nodes $r_{i}$ are the zeros of the $n$th Legendre polynomial $P_{n}(x)$.
Table {\ref{Table2}} lists the nodes and coefficients for n=7 and 8.
More detailed calculations and values of nodes and coefficients can be found in {\cite{Krishnamurthy,Stroud}}.
Such quadrature formulae is called Gauss-Legendre formulae.

The error term of higher-order derivative in quadrature formulae of higher degree is difficult to evaluated or even boundless,
the quadrature formulae \ref{Ori_Gauss} is not recommended to obtain desirable accuracy, although the Gaussian Quadrature is stable.
Instead, we can divide the interval $[a, b]$ into some subintervals $[x_{i}, x_{i+1}]$,
and apply the low-order Gaussian Quadrature to each subinterval.
Then, summing over all the values as the final approximation, we have
\begin{eqnarray}
\int_{a}^{b} f(x) dx = \sum_{i=0}^{m-1}\int_{x_{i}}^{x_{i+1}} f(x) dx,
\end{eqnarray}
where $a = x_{0} < x_{1} < ... < x_{m} = b$ and the subscript $m$ donates the number of the subintervals.

Let the subintervals be of equal size, the composite Gaussian Quadrature can be written as:
\begin{eqnarray}
\label{Gaussian}
\int_{a}^{b} f(x) dx = \frac{h}{2} \sum_{i=1}^{n} c_{i} \left[\sum_{j=0}^{m-1} f(a + \frac{2j+1}{2}h + \frac{h}{2}r_{i})\right],
\end{eqnarray}
where $h = (b-a)/m$ donates the size of the subinterval.

\begin{table}
\centering
\caption{\label{Table2} Nodes and weight coefficients for Gauss-Legendre integration.}
\begin{tabular}{ccc}
  \hline
  n & nodes $r_{n,i}$ & Coefficients $c_{n,i}$ \\ \hline
  7 & $\pm$ 0.9491079123 & 0.1294849662 \\
    & $\pm$ 0.7415311856 & 0.2797053915 \\
    & $\pm$ 0.4058451514 & 0.3818300505 \\
    & 0 & 0.4179591837 \\
  8 & $\pm$ 0.9602898565 & 0.1012285363 \\
    & $\pm$ 0.7966664774 & 0.2223810345 \\
    & $\pm$ 0.5255324099 & 0.3137066459 \\
    & $\pm$ 0.1834346425 & 0.3626837834 \\
  \hline
\end{tabular}
\end{table}

The Gaussian Quadrature formulae can be applied only when $f(x)$ is explicitly known,
so that $f(x)$ can be evaluated for any desired value of $x$.
Naturally, orthogonal polynomial other than the Legendre polynomials also can be used,
such as the Gauss - Chebyshev, Gauss - Jacobi and Gauss - Hermite formulae.

\subsection{Performance of the Two Algorithms}\label{perform}
In the section, we perform the efficiency and accuracy of the two quadrature algorithms with Fortran program.
We create a sample containing about $10^{5}$ redshift data which ranges from 0 to 1100 to present a quantitative comparison.

The Romberg Integration is based on equation {\ref{Romberg}},
and we choose the error tolerance $|R_{n,n} - R_{n,n-1}| < 10^{-6}$ to generate the approximation.
With the fast convergence rate of the iterative, there is no obvious distinctness, including the execution time,
if we set the error tolerance $|R_{n,n} - R_{n,n-1}| < 10^{-4}$ instead.
Different parameters $(m, w_{0}, w_{a})$ are chosen to evaluate the composite Gaussian Quadrature formulae {\ref{Gaussian}},
but we fix the number of nodes $n = 8$.

\begin{figure}[h]
\begin{center}
    \includegraphics[width=0.8\textwidth]{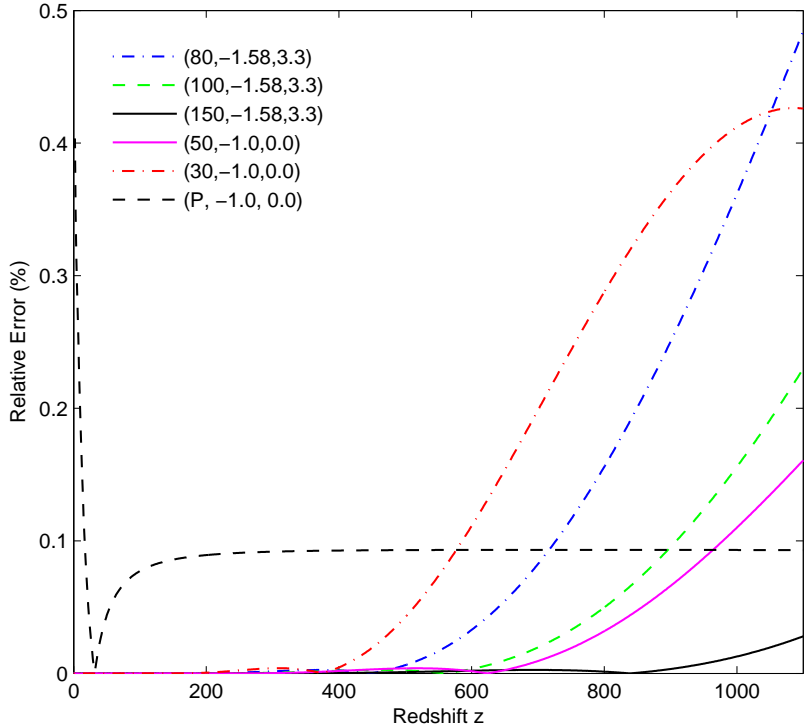}
    \caption{The absolute relative error as a function of $z$ for different parameter choices $(m, w_{0}, w_{a})$.
             The error is no more than $0.5\%$ when $m>80$ for the best fit parameters $w_{0}=-1.58$, $w_{a}=3.3$, $\densm=0.3$.
             The black dashed line, i.e. (P, -1.0, 0.0), denotes the accuracy of the polynomial approximation (see section \ref{lcdm_dl}).}
    \label{fig:err_Gauss}
\end{center}
\end{figure}

Fig {\ref{fig:err_Gauss}} shows the relative error of different parameter choices for Gaussian Quadrature
which illustrates that more subinterval division can improve the accuracy of the algorithm
and the relative error also depends on the choice of $w_{0}$ and $w_{a}$.
However, the error is insensitive to the $m$ for redshift $z<50$.
Actually, the general Gaussian Quadrature based on equation {\ref{Ori_Gauss}}, i.e. $m=1$,
is precise enough to evaluate the integral values in this case.
Therefore, in order to obtain desirable accuracy and efficiency,
we can set a greater number $n$ and suitable $m$ to extend the composite Gaussian Quadrature to a wider redshift distribution.

The main results of the two algorithms are listed in Table {\ref{Table3}}.
Note that the Gaussian Quadrature obviously takes less time than Romberg Integration if the same accuracy is required.
However, extra interpretation about Table \ref{Table3} should be emphasized. The execution time and efficiency
just reflects the relative results of the codes with each other, and depends on the compiler as well
as the different computing environment. More discussions are available in \cite{2011MNRAS.412.2685L} and \cite{2010MNRAS.406..548W}.

\begin{table}
\centering
\caption{\label{Table3} The main results of the Romberg Integration, Gaussian Quadrature and polynomial approximation.
The last two columns show that the iterative of Romberg Integration spends more time, although its accuracy is well under control.}
\begin{tabular}{cccc}
  \hline
   type & parameters & time(s) & maximum error(\%)\\
  \hline
   Romberg Integration & (-1.58, 3.3) & 210.960 & 0.04\\
   Gaussian Quadrature & (80, -1.58, 3.3) & 7.457 & 0.48\\
    & (100, -1.58, 3.3) & 9.266 & 0.23\\
    & (150, -1.58, 3.3) & 13.822 & 0.028\\
    & (200, -1.58, 3.3) & 18.564 & 0.003\\
    & (30, -1.00, 0.0) & 1.888 & 0.43\\
    & (50, -1.00, 0.0) & 3.151 & 0.16\\
  Polynomial approximation & (P, -1.0, 0.0) & 0.016 & 0.41\\
  \hline
\end{tabular}
\end{table}

\section{A Polynomial Approximation to $d_{L}^{\Lambda}$}\label{lcdm_dl}
The general analytical expression of the luminosity distance in the spatially flat $\lcdm$ model is given by equation {\ref{dl_lcdm}}.
Because it is frequently used in practice, many papers have focused on the numerical analysis of the integral equation.
In this section, another polynomial approximation is described for the similar considerations.
Following the notation introduced in our previous work
\cite{2011MNRAS.412.2685L}, substituting $s = \sqrt[3]{(1 - \densm) / \densm}$ and $u = 1 / z' $ into equation {\ref{dl_lcdm}}
yields
\begin{eqnarray}
\frac{d_{L}^{\Lambda}}{c/H_{0}} = \frac{1+z}{\sqrt{s\densm}}\left[{T(s) - T(\frac{s}{1+z})}\right], \nonumber
\end{eqnarray}
where
\begin{eqnarray}
\label{tx}
T (\tau) = \int_{0}^{\tau}\frac{du}{\sqrt{u^4 + u}}.
\end{eqnarray}

However, we note that the behavior of $T(\tau)$ has some deficiencies.  First,
the derivative of $T(\tau)$ becomes singular as $\tau \to 0^{+}$.  Second, the domain
of $T(\tau)$ extends to infinity.  Either one is detrimental to the approximation
using polynomials. Fortunately, such shortages can be eliminated by proper change
of variables. Based on the consideration above, we introduce a mathematical transformation
$x = 1/(\tau+1)$ further to constraint the variation of $x$ within $0 \leq x \leq 1$, which leads to
\begin{eqnarray}
\frac{d_{L}^{\Lambda}}{c/H_{0}} = \frac{1+z}{\sqrt{s\densm}}\left[{f(\frac{1}{s/(1+z)+1}) - f(\frac{1}{s+1})}\right], \nonumber
\end{eqnarray}
where
\begin{eqnarray}
 \label{f}
f(x) = \int_{0}^{x} \frac{du}{\sqrt{(1 - u) \cdot (3u^2 - 3u + 1)}}.
\end{eqnarray}
The main purpose of the section is to obtain the approximate expression of the equation.

Since the integrand intends to infinity as $u \to 1^{-}$, we utilize a roundabout strategy to
achieve the suitable polynomial. By analyzing the integrand, we find that it can be factorized as
above (equation {\ref{f}}), which inspires us to write it as:
\begin{eqnarray}
 \label{appdf}
\hat{f}(x) = \int_{0}^{x} \frac{a_{i}u^{i}du}{\sqrt{1 - u}},
\end{eqnarray}
where $a_{i}u^{i}$ is a polynomial defined as: $a_{i}u^{i} = \sum^{5}_{i=0}a_{i}u^{i}$.

One prominent aspect of the equation {\ref{appdf}} is that it can be integrated analytically.
For quarrying out the six best-fitting free parameters $a_{i}$, we must impose some constraints upon
the equation {\ref{appdf}}. With the modish range of $\densm$ in mind, we just minimize the relative
error between function $\hat{f}$ and $f$ with $0.1< \densm <1$. The restrictions what we adopt are listed as
following:
\begin{eqnarray}
\label{const}
\hat{f}(\frac{1}{3}) = f (\frac{1}{3}), \quad \hat{f}(1) = f (1), \nonumber \\
\hat{f}^{'}(\frac{1}{3}) = f^{'} (\frac{1}{3}), \quad \hat{f}^{'}(1) = f^{'} (1),
\end{eqnarray}
where $1/3$ is approximately equal to the minimum value of $x$ within the considered parameter space,
i.e., $x \in [1/3,1]$.

\begin{figure}[h]
\begin{center}
    \includegraphics[width=0.9\textwidth]{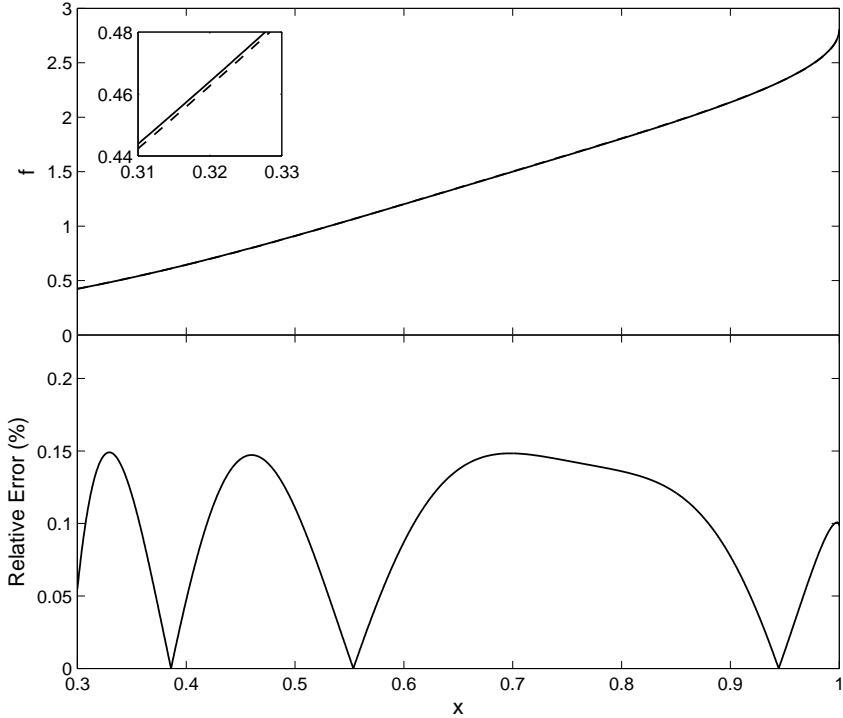}
    \caption{Approximate and analytical function $f$ (\textit{upper panel})
    and the absolute percentage relative error (\textit{lower panel}).
    The solid and dashed lines represent the $\hat{f}$ and $f$ with $0.1 \leq \densm \leq 1$
    in the upper panel, respectively. }
    \label{fig:fg}
\end{center}
\end{figure}

Utilizing equation {\ref{const}}, we can obtain a linear equations which contains just two free parameters.
For instance, we can choice $a_{4}$ and $a_{5}$ as the variables to be determined. In order to derive the total
parameters, we define the relative error as:
\begin{eqnarray}
 \label{err}
e = |\frac{\hat{d_{L}^{\Lambda}}}{d_{L}^{\Lambda}}-1|.
\end{eqnarray}
Just as \cite{1999ApJS..120...49P} has pointed out, the error tends to be
dominated by $z \to 0$, which yields
\begin{eqnarray}
 \label{derr}
e(x)_{max} = |\frac{d\hat{f}}{df}-1|_{z \to 0}.
\end{eqnarray}
The square of $e(x)$, namely $e^{2}(x)$, as a continuous function, is more convenient for us to
acquire the relation between the rest two parameters $a_{4}$ and $a_{5}$. From the mathematical theorem we know
that the first derivative of $e^{2}(x)$, which contains three variables $x$, $a_{4}$ and $a_{5}$,
must be equal to zero strictly if it reaches the local maximum. By solving the three nonlinear equations, the
most appropriate relation between $a_{4}$ and $a_{5}$ can be written as following:
\begin{eqnarray}
\label{a_a}
a_{4} = 12.15722 - 2.92471 \cdot a_{5},
\end{eqnarray}
where $a_{5}$ is still unknown to us. Because we are more interested in the minimum of the $e(x)_{max}$,
substituting equation {\ref{a_a}} into equation {\ref{derr}} and regulating the valve of $a_{5}$ make us
find the most desirable outcome of $e(x)_{max}$ to be about 0.96\% when $a_{5} = -44.63290$.

However, we have emphasized that the implicit requirement of the coincidence of function
values at end points could be unnecessarily strong \cite{2011MNRAS.412.2685L}. If we control the relative
error to the minimum when $\densm$ is taken as the best observational value $0.3$, the final approximation
we construct can be expressed as:
\begin{eqnarray}
\label{appsf}
\hat{f} (x) = \sqrt{1-x} \cdot (8.11507x^5 - 22.69338x^4 + 21.09474x^3 \nonumber \\
        - 6.03039x^2 + 0.32109x - 2.80713) + 2.80713,
\end{eqnarray}
where
\begin{eqnarray}
x = \frac{1}{s/(1 + z) + 1} \nonumber,  \quad s^{3} = \frac{1 - \densm}{\densm}.
\end{eqnarray}
Fig. \ref{fig:fg} shows the results of the fit for the function $f$ (see equation  \ref{f}) and its residual,
which is not more than 0.15\%.

\begin{figure}
\begin{center}
    \includegraphics[width=0.9\textwidth]{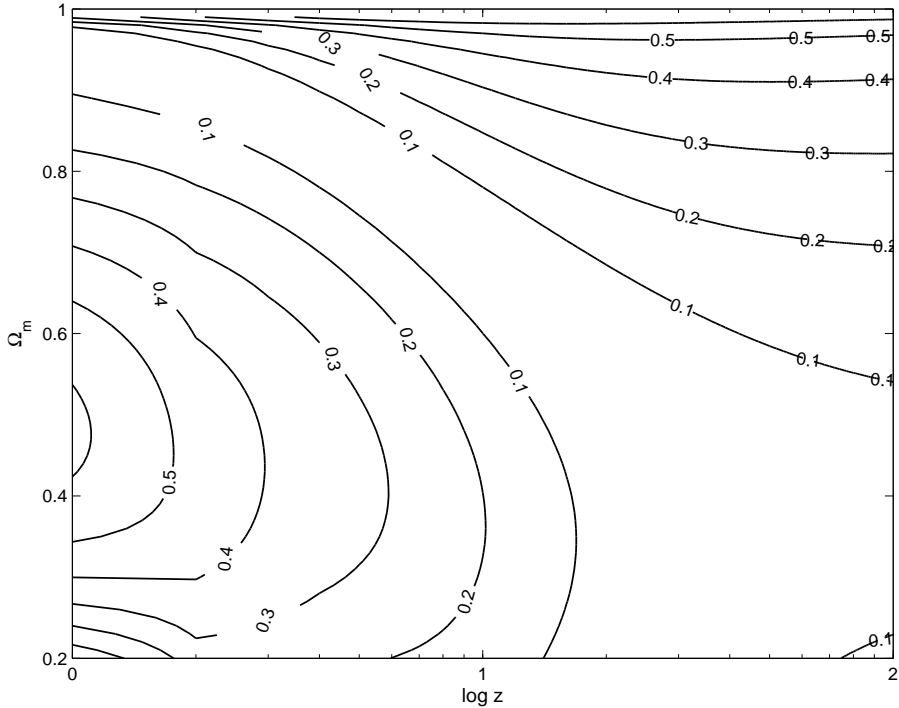}
    \caption{The distribution of the global relative error in $d_{L}$ with different
    $z$ and $\densm$. The maximum error, dominated by the small redshift ($z<0.1$),
    is less than $1\%$.}
    \label{fig:dlerr}
\end{center}
\end{figure}

With the definition of the relative error in equation {\ref{err}}, the distribution of the global relative error
based on the polynomial approximation (equation \ref{appsf}) with various redshift and $\densm$ is plotted in
Fig. \ref{fig:dlerr}. As we can see from the figure, the maximum error, which is less than $1\%$,
tends to be contributed by the small redshift $z$.
Similarly, we compute the running time and accuracy of the approximate luminosity distance based on the same
criterions presented in section \ref{perform}.
From Table {\ref{Table3}} we conclude that the polynomial spends less time than the other two methods,
though its error can't be improved further.
However, Fig. {\ref{fig:err_Gauss}} shows that the accuracy of the polynomial at low redshift is inferior to
the composite Gaussian Quadrature. And as the increase of redshift $z$, the relative error tends to be around $0.09\%$.

Because approximating the integrand directly can decrease the error of the target-integral,
one promising extension of our method is to the linear growth factor $\delta(a)$ \cite{2007APh....28..481L, 2008MNRAS.387.1126B}
contained only $\densm$ and $\densl$. The general form of $\delta(a)$ in $\lcdm$ universe can be written as
\begin{eqnarray}
\label{grof}
\delta(a) = \frac{5\densm}{2}\frac{H(a)}{H_{0}}\int_{0}^{a}\frac{da'}{[a'H(a')/H_{0}]^{3}},
\end{eqnarray}
where $a$ and $H(a)$ represent the scale factor and Hubble parameter, respectively.
Kasai \cite{2010arXiv1012.2670A} has developed an effective recipe for evaluating
it recently. However, a more compact form may be helpful to research the matter density perturbation with numerical
simulation efficiently, and as a candidate our idea behind the method may be useful.

\section{Conclusions}\label{conclu}
Two different quadrature algorithms are presented to evaluate the integral involved in luminosity distance in CPL parametrization model.
Through the comparison of the efficiency and accuracy, the composite Gaussian Quadrature is more promising to apply into the evaluation of
such complex integral. Because of the generalization of the algorithms,
we can also extend them to other parametric models of dark energy,
for instance the two-index parameterizations by Huterer et al.{\cite{Huterer}},
and the four-index parameterizations by Hannestad et al.{\cite{Hannestad}}.

Additionally, because general Gaussian Quadrature is precise enough for refshift $z<50$, it is unnecessary to divide the interval $[a, b]$
to be of equal size. The adaptive integration combined with the composite Gaussian Quadrature may be more helpful.

\acknowledgments
We are very grateful to the referee for many valuable comments that greatly
improved the paper. This work was supported by the National Science Foundation of China (Grants No. 11173006),
the Ministry of Science and Technology National Basic Science program (project 973) under grant No. 2012CB821804,
and the Fundamental Research Funds for the Central Universities.


\end{document}